\documentclass{article}
\usepackage{amsmath,latexsym,amssymb,color}
\title{Algebraic aspects of quantum indiscernibility}
\author{\\D\'ecio Krause\\
UFSC -- Department of Philosophy\\
deciokrause@gmail.com
\\
\\
H\'ercules de Araujo Feitosa\\
UNESP -- FC -- Department of Mathematics\\
haf@fc.unesp.br}

\newtheorem{thm}{Theorem}[section]

\newtheorem{dfn}{Definition}[section]

\newcommand{\Proof}{\noindent\textit{Proof:} \,}
\newcommand{\cqd}{{\rule{.60ex}{1.7ex}}} %{\hfill{$\Box$}}

\newcommand{\igual}{=_{\small\mathsf{def}}}

\newcommand\oA{\overline{A}}
\newcommand\oB{\overline{B}}
\newcommand\oC{\overline{C}}

\newcommand\ove{\overline}
\newcommand{\Q}{$\mathfrak{Q}$}

\newcommand{\Lq}{\mathcal{L}_q}
\newcommand{\Al}{\mathfrak{L}}
\newcommand{\lra}{\leftrightarrow}

\begin{document}
\maketitle

\begin{abstract}
\noindent Quasi-set theory was proposed as a mathematical context
to investigate collections of indistinguishable objects. After
presenting an outline of this theory, we define an algebra that
has most of the standard properties of an orthocomplete
orthomodular lattice, which is the lattice of the closed subspaces
of a Hilbert space. We call the mathematical structure so obtained
$\mathfrak{I}$-lattice. After discussing, in a preliminary form,
some aspects of such a structure, we indicate the next problem of
axiomatizing the corresponding logic, that is, a logic which has
$\mathfrak{I}$-lattices as its algebraic models. We suggest that
the intuitions that the `logic of quantum mechanics' would be not classical logic (with its Boolean algebra), is consonant with
the idea of considering indistinguishability right from the start,
that is, as a primitive concept. In other words, indiscernibility
seems to lead `directly' to $\mathfrak{I}$-lattices. In the first
sections, we present the main motivations and a `classical'
situation which mirrors that one we focus on the last part of the
paper. This paper is our first study of the algebraic structure of
indiscernibility within quasi-set theory.
\end{abstract}

%-----------------------------------------------------------------
\renewenvironment{enumerate}{\begin{list}{}{\rm \labelwidth 0mm
\leftmargin 5mm}} {\end{list}}

\section{Introduction}
Indiscernibility is a typical concept of quantum physics, and some
facts implied by indiscernibility, as the properties of a
Bose-Einstein condensate, have no parallel in classical physics.
Without considering that quanta are indiscernible, no explanation
of colors would be done, no vindication to the periodic table of
elements would result, and, among other things, Planck would not
arrive to his formula for the black body radiation. Some authors
have sustained that quantum indiscernibility results from the
raise of quantum ``statistics" (really, ways of counting), while
others think that they can \emph{explain} quantum statistics
without presupposing indiscernibility, but at the expenses of
rejecting equiprobability.\footnote{This is the particular van
Fraassen's view; for instance, he supposes two particles 1 and 2
in two possible states $A$ and $B$, and the possible cases are:
(i) 1 and 2 in $A$; (ii) 1 and 2 in $B$, both cases with
probability 1/3 each; (iii) 1 in $A$ and 2 in $B$; (iv) 1 in $B$
and 2 in $A$, both with probability 1/6. According to this author,
this way we can arrive at Bose-Einstein statistics \cite{van98}.
But the problem is that situations (iii) and (iv) need to be
distinguished from one another, and if the involved quanta are
indiscernible, this can be done only either by the assumption of
some kind of hidden variable or by some form of \emph{substratum},
and we know that both possibilities conduce to well known
problems.} That discussion is still alive, and we have much to do
in the philosophical, epistemological, logical, and on ontological
aspects of quantum indiscernibility, mainly if we agree (with
Arthur Fine) that philosophy of science should be engaged with
on-going science (\emph{apud} \cite{fre98}). This leads us
directly to the quantum field theories, and perhaps more, to
string theories and to quantum gravitation. Acknowledging this
naturalistic claim, we shall be here quite modest in discussing
some algebraic aspects of a mathematical theory which was
conceived to deal with indistinguishable objects, the
\emph{quasi-set theory}. Without revising all the details of such
a theory (to which we refer to Chapter 7 of \cite{frekra06}), we
shall keep the paper self-contained so that the reader can
understand the basic ideas, although sometimes in the intuitive
sense, and only the really necessary concepts and postulates are
mentioned.

It should be recalled that indiscernibility enters in the standard
quantum formalism by means of symmetry postulates. The relevant
functions for systems of many quanta ought to be either
symmetrical or anti-symmetrical, and this assumption makes the
expectation values to assume the same values before and after a
permutation of indiscernible elements. Thus, physicists (and
philosophers accept that) say that ``the individuality was lost,''
as if there would be something to lose. In this work, we enlarge
our research program of providing a mathematical basis for quantum
theory that takes indiscernibility ``right from the start", as
claimed by Heinz Post \cite{pos63} (see \cite{frekra06}), with the
algebraic discussion of indiscernibility. All the considerations
are performed within quasi-set theory, which we revise in its main
ideas below.

%-----------------------------------------------
\section{Quasi-sets}\label{qsets}
Quasi-set theory, denoted by \Q, was conceived to handle
collections of indistinguishable objects, and was motivated by
some considerations taken from quantum physics, mainly in what
respects Schr\"odinger's idea that the concept of identity cannot
be applied to elementary particles \cite[p.\ 17-18]{sch52}. Of
course, the theory can be developed independently of quantum
mechanics, but here we shall have this motivation always in mind.
Our way of dealing with indistinguishability is to assume that
expressions like $x=y$ are not well formed in all situations that
involve $x$ and $y$. We express that by saying that the concept of
identity does not apply to the entities denoted by $x$ and $y$ in
these situations. Here, quantum objects do not mean necessarily
\emph{particles}, but ought to be thought as representing the
basic objects of quantum theories, which although differ form one
theory to another (\cite[Chapter 6]{fal07}), have some common
characteristics, as those related to indiscernibility (with the
exception of some hidden variable theories, like Bohm's, which
will be not discussed here).\footnote{Since such theories present
difficulties due to results like Kochen-Specker theorem and Bell's
inequalities, so as due to the fact that apparently they cannot be
extended to quantum field theories, we shall leave them outside of
our discussion.} Due to the lack of sense in applying the concept
of identity to certain elements, informally, a quasi-set (qset),
that is, a collection of such objects, may be such that its
elements cannot be identified by names, counted, ordered, although
there is a sense in saying that these collections have a cardinal.
This concept of cardinal is not defined by means of ordinals, as
usual -- see below. But we aim at to keep standard mathematics
intact,\footnote{So respecting the quite strange rule what
Birkhoff and von Neumann call ``Henkel's principle of the
`perseverance of formal laws' ", explained by R\'edei as ``a
methodological principle that is supposed to regulate mathematical
generalizations by insisting on preserving certain laws in the
generalization" \cite{red07}; of course we are `preserving' all
standard mathematics built in ZFC.} so the theory is developed in
a way that ZFU (and hence ZF, perhaps with the axiom of choice,
ZFC) is a subtheory of \Q. In other words, there is a ``copy" of
ZFU into \Q, that is, this theory is constructed so that it
extends standard Zermelo-Fraenkel with \textit{Urelemente} (ZFU).
In this case, standard sets of ZFU can be viewed as particular
qsets, that is, there are qsets that have all the properties of
the sets of ZFU, and we call then \Q-sets. The objects in \Q\ that
correspond to the \textit{Urelemente} of ZFU are termed $M$-atoms.
But quasi-set theory encompasses another kind of
\textit{Urelemente}, the $m$-atoms, to which the standard theory
of identity does not apply. More especially, expressions like $x=
y$ are not well formed when $m$-atoms are involved.

When \Q\ is used in connection with quantum physics, these
$m$-atoms are thought as representing quantum objects (henceforth,
q-objects), and not necessarily they are `particles', as mentioned
above; waves or perhaps even strings (and whatever `objects'
sharing the property of indistinguishability of pointing
elementary particles) can be also be values of the variables of
\Q. The lack of the concept of identity for the $m$-atoms make
them \emph{non-individuals} in a sense, and it is mainly (but not
only) to deal with collections of $m$-atoms that the theory was
conceived. So, \Q\ is a theory of generalized collections of
objects, involving non-individuals. For details about \Q\ and
about its historical motivations, see \cite[p.\ 119]{cos80},
\cite{dalgiukra98}, \cite[Chapter\ 7]{frekra06}, \cite{kra92}, and
\cite{krasansar05}.

In \Q, the so called `pure' qsets have only q-objects as elements
(although these elements may be not always indistinguishable from
one another), and to them it is assumed that the usual notion of
identity cannot be applied, that is, $x=y$, so as its negation, $x
\not= y$, are not a well formed formulas if $x$ and $y$ stand for
q-objects. Notwithstanding, there is a primitive relation
$\equiv$ of indistinguishability having the properties of an
equivalence relation, and a concept of \textit{extensional
identity}, not holding among $m$-atoms, is defined and has the
properties of standard identity of classical set theories.  Since
the elements of a qset may have properties (and satisfy certain
formulas), they can be regarded as \textit{indistinguishable}
without turning to be \textit{identical} (being
\textit{the same} object), that is, $x \equiv y \nRightarrow
x = y$.

Since the relation of equality, and the concept of
identity, does not apply to $m$-atoms, they can also be thought
as entities devoid of individuality. We further remark that if the
`property' $x=x$ (to be identical to itself, or
\textit{self-identity}, which can be defined for an object $a$ as
$I_a(x) \igual x = a$) is included as one of the properties of the
considered objects, then the so called Principle of the Identity
of Indiscernibles (PII) in the form $\forall F (F(x)
\leftrightarrow F(y)) \to x=y$ is a theorem of classical second-order
logic, and hence there can not be indiscernible but not
identical entities (in particular, non-individuals). Thus, if
self-identity is  linked to the concept of non-individual, and if
quantum objects are to be considered as such, these entities fail
to be self-identical, and a logical framework to accommodate them
is in order (see \cite{frekra06} for further argumentation).

We have already discussed at length in the references given above
(so as in other works) the motivations to build the quasi-set
theory, and we shall not return to these points here,\footnote{But
see  \cite{coskra94}, \cite{coskra97}, \cite{coskra07},
\cite{frekra06}, \cite{kra96}, \cite{krasansar05}.} but before to
continue we would like to make some few remarks on a common
misunderstanding about PII and quantum physics. People generally
think that spatio-temporal location is a sufficient condition for
individuality. Thus, an electron in the South Pole and another one
in the North Pole \emph{are} discernible, hence \emph{distinct
individuals}, so that we can call ``Peter" one of them and ``Paul"
the another one. Leibniz himself prevented us about this claim
(yet not directly about quantum objects of course), by saying that
``it is not possible for two things to differ from one another in
respect to place and time alone, but that is always necessary that
there shall be some other internal difference" \cite{lei95}.
Leaving aside a possible interpretation for the word `internal',
we recall that even in quantum physics, where fermions obey the
Pauli Exclusion Principle, which says that two fermions (yes, they
`count' as more than one) can not have all their quantum numbers
(or `properties') in common, two electrons (which are fermions),
one in the South Pole and another one in the North Pole, \emph{are
not individuals in the standard sense}.\footnote{Without aiming at
to extend the discussion on this topic here (see \cite{frekra06}),
like an individual we understand an object that obeys the
classical theory of identity of classical logic (extensional set
theory included).} In fact, we can say that the electron in the
South Pole is described by the wave function $\psi_S$, while the
another one is described by $\psi_N$ (words like `another' in the
preceding phrase are just ways of speech). But the joint system
is, in a simplified form, given by $\psi_{SN} = \psi_S - \psi_N$
(the function must be anti-symmetric in the case of fermions, that
is, $\psi_{SN} = - \psi_{NS}$), a superposition of the two first
wave functions, and this last function cannot be factorized.
Furthermore, in the quantum formalism, the important thing is the
square of the wave function, which gives the joint probability
density; in the present case, we have $|| \psi_{SN} ||^2 = ||
\psi_S ||^2 + || \psi_N ||^2 - 2\mathrm{Re} \psi_S \psi_N$. This
last term, called `the interference term,' can not be dispensed
with, and says that nothing, not even \emph{in mente Dei}, can
tell us which is the particular electron in the South Pole (and
the same happens for the North Pole), that is, we never will know
who is Peter and who is Paul, and in the limits of quantum
mechanics, this is not a matter of epistemological ignorance, but
it is rather an ontological question. As far as quantum physics is
concerned with its main interpretations, they  seem to be really
and truly objects without identity.

In the next sections, we shall discuss from an algebraic point of
view some issues of non-individuality. It should be interesting to
recall that the `qset'-operations of intersection ($\cap$), union
($\cup$), difference ($-$) work similarly in \Q\ as the standard
ones in usual set theories.

%----------------------------------------------
\section{Algebraic aspects: the lattice of indiscernibility}

Quantum logic was born with Birkhoff and von Neumann's paper from
1936 \cite{birvon36}. Today it consists in a wide field of
knowledge, having widespread to domains never thought by the two
celebrated forerunners. For a look on the state of the art, see
\cite{dalgiugre04}. The main idea is that the typical algebraic
structures arising from the mathematical formalism of quantum
mechanics is not a Boolean algebra, but an orthocomplete
($\sigma$-orthocomplete in the general case \cite[p.\
39]{dalgiugre04}) orthomodular lattice. We shall see below that in
quasi-set theory, by considering indiscernibility right from the
start, a similar structure `naturally' arises. Let us provide the
details before ending with some comments and conclusions.

Now we need of the concept of Tarski's system and topological
space.

\begin{dfn} [Tarski's Space] A Tarski's Space is a pair $(E, ^-)$
where $E$ is a non empty set and $^-$ be a function $^-:
\mathcal{P}(E) \rightarrow \mathcal{P}(E)$, called the Tarski's
consequence operator, such that: (i) $A \subseteq \overline{A}$;
(ii) $A \subseteq B \Longrightarrow \overline{A} \subseteq
\overline{B}$; (iii) $\overline{\overline{A}} \subseteq
\overline{A}$.
\end{dfn}\label{TK-space}

\begin{thm} In any Tarski's Space $(E, ^-)$ we have that:
\begin{enumerate} \item (i) $\overline{\overline{A}} = \overline{A}$;
\item (ii) $\oA \cup \oB \subseteq \ove{A \cup B}$; \item (iii)
$\ove{A \cap B} \subseteq \oA \cap \oB$; \item (iv) $\ove{\oA \cup
\oB} = \ove{A \cup B}$; \item (v) $\oA \cap \oB = \ove{\oA \cap
\oB}$.
\end{enumerate} \end{thm} \Proof  See [13]. \cqd

\begin{dfn} [Closed and open sets] In a Tarski's Space $(E, ^-)$,
a subset $A \subseteq E$ is closed when $\overline{\overline{A}} =
\overline{A}$ and $A$ is open when its complement relative to $E$,
denoted by $A^C$, is closed.
\end{dfn}\label{closed and open}

\begin{dfn} [Closure and interior] Given $A \subseteq E$ in $(E,
^-)$, the set $\overline{A}$ is the closure of $A$ and the set
$\AA = (\overline{A^C})^C$ is the interior of $A$.
\end{dfn}\label{closure and interior}

\begin{thm} In any Tarski's Space $(E, ^-)$ it follows that:  $\AA
\subseteq A \subseteq \overline{A}$. \end{thm}

\begin{dfn} [Topological Space] A Topological Space is a pair $(E, ^-)$
where $E$ is a non empty set and $^-$ be a function $^-:
\mathcal{P}(E) \rightarrow \mathcal{P}(E)$, such that:  (i) -
(iii) of Definition 3.1 hold plus (iv) $\overline{A \cup B} =
\overline{A} \cup \overline{B}$; (v) $\overline{\emptyset} =
\emptyset$.
\end{dfn}\label{Topological}

Hereafter, we shall be working in the theory \Q, and use the
equality symbol $=$ to stand for the extensional equality  of \Q.
Intuitively speaking,  $x = y$ holds when $x$ and $y$ are both
qsets and have the same elements (in the sense that an object
belongs to $x$ iff it belongs to $y$) or they are both $M$-objects
and belong to the same qsets. It can be proven that $=$ has all
properties of standard identity of first-order ZFC. Qsets which
may have $m$-atoms as elements are written (in the metalanguage)
with square brackets ``[" and ``]", and \Q-sets (qsets whose
transitive closure have no $m$-atoms) with the usual curly braces
``\{" and ``\}".

We start with the concept of cloud that will to point to the
algebraic aspects whose we are involved.

\begin{dfn} [Cloud] Let $U$ be a non empty qset and $A$ be
a subqset of $U$. The \emph{cloud} of $A$ is the qset
$$\overline{A} \igual [y \in U : \exists x (x \in A \wedge y
\equiv x)].$$
\end{dfn}\label{closure}

Intuitively speaking, $\oA$ is the qset of the elements of $U$
(the universe) which are indistinguishable from the elements of
$A$. If $A$ is a \Q-set, that is, a copy of a set of ZFU, then of
course the only indistinguishable of a certain $x$ is $x$ itself,
thus $\oA = A$.

\begin{thm} The application\footnote{In \Q, the concept of function
must be generalized, for if there are $m$-atoms involved, a
mapping in general does not distinguish between arguments and
values. Thus we use the notion of q-function, which leads
indistinguishable objects into indistinguishable objects, and
which reduces to standard functions when there are no $m$-atoms
involved. Thus, from the formal point of view, the defined mapping
may associate to $A$ whatever qset from a collection of
indistinguishable qsets. But this does not matter. As in quantum
physics, it is not the extension of the collections which are
important; informally saying, \emph{any} elementary particle of a
certain kind serves for all purposes involving it. This is the
principle of the invariance of permutations.} that associates to
every subqset of $U$ its cloud is a Tarski's operator and $(U,
^-)$ is Tarski's Space.
\end{thm} \Proof  (i) $A \subseteq \oA$: Let $t \in A$. Then,
by the reflexivity of $\equiv$, we have $t \equiv t$, hence $t \in
\oA$. (ii) $A \subseteq B \Rightarrow \oA\ \subseteq \oB$: Let $A
\subseteq B$, and let $t \in \oA$. Then there exists $x \in A$
such that $t \equiv x$. Since $x \in B$, then $t \in \oB$. (iii)
$\overline{\oA} \subseteq \oA$: Let $t \in \overline{\oA}$. Then
there exists $x \in \oA$ such that $t \equiv x$. But then there
exists $y \in A$ such that $x \equiv y$. By the transitivity of
$\equiv$, we have $t \equiv y$, hence $t \in \oA$. \cqd

{}From now on, we shall suppose that $U$ is closed, that is, it
contains all the indistinguishable objects of its elements. Some
interpretations linked to physical situations are possible. For
instance, $\oA$ can be thought as the region where the wave
function $A$ of a certain physical system is different from zero.
Another possible interpretation is to suppose that the clouds
describe the systems plus the cloud of virtual particles that
accompany those of the considered system. But in this paper we
shall be not considering these motivations, but just to explore
its algebraic aspects.

It is immediate to prove the following theorem:

\begin{thm}\label{cloud_top} $(U, ^-)$ is a topological space.
\end{thm}
\Proof \begin{enumerate} \item (i) $\oA \cup \oB = \ove{A \cup
B}$: $A \subseteq A \cup B$, so $\oA\ \subseteq \ove{A \cup B}$.
In the same way, $B \subseteq A \cup B$ and $\oB\ \subseteq \ove{A
\cup B}$. Thus $\oA \cup \oB \subseteq \ove{A \cup B}$.
Conversely, suppose $t \in \ove{A \cup B}$, then there is $x \in A
\cup B$ such that $t \equiv x$. So there is $t \equiv x$ such that
$x \in A$ or $x \in B$. In this way $t \in \oA$ or $t \in \oB$ and
therefore $t \in \oA \cup \oB$. \item (ii) It follows immediately
from the definition of cloud that $\overline{\emptyset} =
\emptyset$. \cqd
\end{enumerate}

\vspace{2mm}
The next definition introduces the lattice operations on subqsets
of a qset $U$, the universe.

\begin{dfn}[$\mathfrak{I}$-lattice operations] \label{operations}
Let $A, B \subseteq U$. Then:
\begin{enumerate} \item ($\sqcap$) $A \sqcap B \igual \ove{A \cap B}$;
\item ($\sqcup$) $A \sqcup B \igual \oA \cup \oB$;
\item (0) $\mathbf{0} \igual \emptyset$;
\item (1) $\mathbf{1} \igual U$. \end{enumerate} \end{dfn}

We note that even if $A \cap B = \emptyset$, may be that $\oA \cap
\oB \not= \emptyset$.

\begin{thm}\label{pp} For any $A, B \in \mathcal{P}(U)$:
    \begin{enumerate}
    \item  (i) $A \sqcap B \subseteq \ove{(\oA \cap \oB)}$;
    \item (ii) $A \sqcap B \subseteq A \sqcup B$;
    \item (iii) If $A$ and $B$ are closed, $A \cup B$ and $A \cap B$
    are closed,     and $A \sqcap B = \oA \cap \oB$.
    \end{enumerate}
\end{thm}
\Proof  \begin{enumerate} \item (i) Immediate, since $\ove{A \cap
B} \subseteq \oA \cap \oB$ (Theorem 3.1 (iii)); \item (ii) $A
\sqcap B$ = $\ove{A \cap B} \subseteq$  (Theorem 3.1 (iii))
$\subseteq \oA \cap \oB \subseteq \oA \cup \oB$ = $A \sqcup B$;
\item  (iii) If $\oA = A$ and $\oB = B$, then $A \cup B$ = $\oA
\cup \oB$ = $\ove{A \cup B}$ (Theorem 3.4 (i)). Furthermore, the
same hypothesis entails that $\ove{A \cap B}$ = $\ove{\oA \cap
\oB}$ = (Theorem 3.1 (v)) $\oA \cap \oB$ = $A \cap B$. Finally,
since $\oA \cap \oB$ = $\ove{(\oA \cap \oB)}$ = $\ove{A \cap B}$ =
$A \sqcap B$ (Theorem 3.1 (v) and the hypothesis). \cqd
\end{enumerate}

\begin{thm} Let $\mathcal{C}$ be the qset of all closed subqsets of $U$.
Then the structure $\mathfrak{C}$ = $\langle \mathcal{C},
\sqcap, \sqcup, \mathbf{0}, \mathbf{1} \rangle$ is a lattice with
0 and 1. But, if we consider also the sub-qsets of $U$ that are
not closed, then some of the properties of such a structure do not
hold, as we emphasize in the proof below.
\end{thm}
\Proof  In this case, for every $A \subseteq U$ holds $\oA = A$.
Firstly, it is immediate to see that if $U \not= \emptyset$, then
$\mathcal{P}(U) \not= \emptyset$. Furthermore, we can prove that
$A \cap (B \cap C) = (A \cap B) \cap C$ and $A \cup (B \cup C) =
(A \cup B) \cup C$ for closed qsets.
\begin{enumerate}
    \item (a) Idempotency (restricted to closed qsets): $A \sqcap A$ =
    $\ove{A \cap A} = \oA$ ($= A$ when $A$ is closed). Also, $A \sqcup A$ =
    $\oA \cup \oA = \oA$ ($= A$ when $A$ is closed). If $A$ is not closed,
    then $A \sqcap A = \oA \neq A$ and $A \sqcup A = \oA \neq A$;
    \item (b) Commutativity (unrestricted): $A \sqcap B$ = $\ove{A \cap B}$
    = $\ove{B \cap A} = B \sqcap A$. In the same way, $A \sqcup B$ =
    $\oA \cup \oB$ = $\oB \cup \oA$ = $B \sqcup A$;
    \item (c) Associativity (unrestricted): (we shall be using items (iii)
    and (iv) of Theorem 3.1 without
    mentioning):
        \begin{enumerate}
        \item (i) $A \sqcap (B \sqcap C)$ = $A \sqcap (\oB \cap \oC)$ =
        $\oA \cap \ove{(\oB \cap \oC)}$ = $\oA \cap (\oB \cap \oC)$ =
        $(\oA \cap \oB) \cap \oC$ = $\ove{(\oA \cap \oB)} \cap \oC$ =
        $\ove{(\oA \cap \oB)} \sqcap C$ = $(A \sqcap B) \sqcap C$;
        \item (ii) $A \sqcup (B \sqcup C) = A \sqcup (\oB \cup \oC) =
        \oA \cup \ove{(\oB \cup \oC)} = \ove{\oA} \cup \ove{(\oB \cup \oC)}$
        = (Theorem 3.1 (i)) $\ove{\oA \cup (\oB \cup \oC)} =
        \ove{(\oA \cup \oB) \cup \oC}$ =  (Theorem 3.1 (i)) $\ove{(\oA \cup \oB)} \cup \ove{\oC}$
        = $\ove{(\oA \cup \oB)} \cup \oC$ = $(\oA \cup \oB) \sqcup C$ = $(A \sqcup B)
        \sqcup C$;
        \end{enumerate}
    \item (d) Absorption (restricted):
        \begin{enumerate}
        \item (i) $A \sqcap (A \sqcup B)$ = $A \sqcap (\oA \cup \oB)$ = $\ove{A \cap (\oA \cup \oB)}$.
        But $A \subseteq \oA$, so $A \subseteq \oA \cup \oB$, then $\ove{A \cap (\oA \cup \oB)}$ =
        $\oA$ \, $(= A$ when $A$ is a closed qset);
        \item (ii) $A \sqcup (A \sqcap B)$ = $A \sqcup \ove{(A \cap B)}$ =
        $\oA \cup \ove{\ove{(A \cap B)}}$ =
        $\oA \cup \ove{(A \cap B)} = \oA$, for $A \cap B \subseteq A \subseteq \oA$ $
        (= A$ when $A$ is a closed qset);
        \end{enumerate}
    \item (e) The properties of $\mathbf{0}$ and $\mathbf{1}$:
        \begin{enumerate}
        \item (i) $\mathbf{0} \sqcap A = \ove{\emptyset \cap  A} = \ove{\emptyset} = \emptyset =
        \mathbf{0}$;
        \item (ii) $\mathbf{0} \sqcup A = \ove{\emptyset} \cup \oA$ = $\emptyset \cup \oA$ =
        $\oA \,$ (= $A$ when $A$ is a closed qset);
        \item (iii) $A \sqcap \mathbf{1} = \ove{A \cap U} = \oA \, (= A$ when $A$ is a closed qset);
        \item (iv) $A \sqcup \mathbf{1} = \oA \cup \ove{U} =
        \ove{U} = U = \mathbf{1}$ (recall our initial hypothesis that $U$ is closed). \cqd
        \end{enumerate}
\end{enumerate}

\begin{thm} The lattice $\mathfrak{C}$ of the closed qsets of $U$ is distributive.
\end{thm}
\Proof We shall emphasize those passages which make use of the hypothesis
that the qsets are closed.
\begin{enumerate}
\item (i) $A \sqcup (B \sqcap C)$ = $A \sqcup \ove{(B \cap C)}$ =
$\oA \cup \ove{\ove{(B \cap C)}}$ = [Th. 3.1 (ii)] $\ove{A \cup
\ove{(B \cap C)}}$ = $ \ove{\oA \cup \ove{(B \cap C)}}$ = [Th. 3.1
(iv)] $\ove{A \cup (B \cap C)}$ = $\ove{(A \cup B) \cap (A \cup
C)}$ = [Th. 3.1 (iii) and because $A \cup B$ and $A \cup C$ are
both closed, for otherwise the equality does not hold] $\ove{(A
\cup B)} \cap \ove{(A \cup C)}$ = [Th. 3.1 (ii)] $(\oA \cup \oB)
\cap (\oA \cup \oC)$ = $\ove{(\oA \cup \oB)} \cap \ove{(\oA \cup
\oC)}$ = [Th. 3.1 (v)] $\ove{(\oA \cup \oB) \cap (\oA \cup \oC)}$
= $(\oA \cup \oB) \sqcap (\oA \cup \oC)$ = $(A \sqcup B) \sqcap (A
\sqcup C)$; \item (ii) $(A \sqcap B) \sqcup (A \sqcap C) = \ove{(A
\cap B)} \sqcup \ove{(A \cap C)} = \ove{\ove{(A \cap B)}} \cup
\ove{\ove{(A \cap C)}}$ = [Th. 3.1 (ii)] $\ove{\ove{(A \cap B)}
\cup \ove{(A \cap C)}}$ = [for $A \cap B$ and $A \cap C$ are
closed] $\ove{(A \cap B) \cup (A \cap C)}$ = $\ove{A \cap (B \cup
C)}$ = (for closed qsets) $\ove{\oA \cap \ove{(B \cup C)}}$ = [Th.
3.1 (i)] $\ove{\oA \cap (\oB \cup \oC)}$ = [since $A$ is closed]
$\ove{A \cap (B \sqcup C)} = A \sqcap (B \sqcup C)$. \cqd
\end{enumerate}

\vspace{2mm} This result is not surprising, for we are dealing
with set theoretical operations which, defined on the closed qsets
of $U$, act as the usual set theoretical properties on standard
sets. But if we consider \emph{all} qsets in $U$ and not only the
closed ones, the distributive laws do not hold, as we can see from
the above proof, which makes essential use of the fact that the
involved qsets are closed (without such an hypothesis, the proof
does not follow). Since the corresponding structure $\mathfrak{I}
= \langle \mathcal{P}(U), \sqcap, \sqcup, \mathbf{0}, \mathbf{1}
\rangle$ has similarities with a lattice with 0 and 1, we propose
to call it \emph{the lattice of indiscernibility}, or just
$\mathfrak{I}$-lattice for short. Other distinctive
characteristics of this ``quasi-lattice" are obtained when we
introduce other operations similar to those of order and
involution, or generalized complement \cite[p.\ 11]{dalgiugre04}.
At Section 4, we sum up the main properties of an
$\mathfrak{I}$-lattice.

\begin{dfn} [$\mathfrak{I}$-order] $A \leq B \igual A \sqcup B =
\oB$.
\end{dfn}

\begin{thm}  The order relation obeys the following properties: \end{thm}
\begin{enumerate}
\item (i) $A \leq A$ and $A \leq \oA$; \item (ii) $A \leq B$ and
$B \leq A \Rightarrow \oA = \oB$ \, (and $A = B$ if they are both
closed); \item (iii) $A \leq B$ and $B \leq C \Rightarrow A \leq
C$ \item (iv) $A \sqcap B \leq A$, and $A \sqcap B \leq B$; \item
(v) $C \leq A$ and $C \leq B \Rightarrow C \leq A \sqcap B$ \item
(vi) $A \leq A \sqcup B$, $B \leq A \sqcup B$; \item (vii) $A \leq
C$ and $B \leq C \Rightarrow A \sqcup B \leq C$; \item (viii)
$\textbf{0} \leq A$, and $A \leq \textbf{1}$ (recall that
$\mathbf{1} = U$ is closed); \item (ix) $A \leq B \Rightarrow A
\sqcap B = \oA$.
\end{enumerate}

\Proof  \begin{enumerate} \item (i) $A \sqcup A = \oA \cup \oA =
\oA$, so $A \leq A$; and $A \sqcup \oA = \oA \cup \ove{\oA} = \oA
\cup \oA = \oA$, so $A \leq \oA$; \item (ii) $A \leq B \Rightarrow
\oA \cup \oB = \oB$, while $B \leq A \Rightarrow \oB \cup \oA =
\oA$, since $\oA  \cup \oB = \oB \cup \oA$,  then $\oA = \oB$ ($A
= B$ for closed qsets); \item (iii) If $A \leq B$ and $B \leq C$,
then $\oA \cup \oB = \oB$ and $\oB \cup \oC = \oC$, therefore $\oA
\cup \oC = \oA \cup (\oB \cup \oC) = (\oA \cup \oB) \cup \oC = \oB
\cup \oC = \oC$, that is, $A \leq C$; \item (iv) $A \sqcap B \leq
A$ iff $\ove{(A \cap B)} \cup \oA = \oA$. But, by Theorem\ 3.4
(i), $\ove{(A \cap B)} \cup \oA = \ove{(A \cap B) \cup A} = \oA$.
Equivalently, $A \sqcap B \leq B$ iff $\ove{(\oA \cap \oB)} \cup
\ove{\oB} = \oB$. But, by Theorem 3.4 (i), $\ove{(\oA \cap \oB)}
\cup \ove{\oB} = \ove{(\oA \cap \oB) \cup \oB} = \ove{\oB} = \oB$;
\item (v) $C \leq A \sqcap B$ iff $\oC \cup \ove{(A \sqcap B)} =
\ove{(A \sqcap B)}$ = [Theorem\ 3.1 (v)] $\oA \cap\oB$. But the
hypothesis tells us that $\oC \cup \oA = \oA$ and $\oC \cup \oB =
\oB$, hence $\oC \subseteq \oA$ and $\oC \subseteq \oB$, that is,
$\oC \subseteq (\oA \cap \oB)$. So, $\oC \cup (\oA \cap \oB) = \oA
\cap \oB$, that is, $C \leq A \sqcap B$; \item (vi) $A \leq A
\sqcup B$ iff $A \sqcup (A \sqcup B) = \ove{A \sqcup B} =
\ove{(\oA \cup \oB)} = \ove{A \cup B}$ by Theorem 3.4 (i). But  $A
\sqcup (A \sqcup B) = A \sqcup (\oA \cup \oB) = \oA \cup \ove{(\oA
\cup \oB)} = \oA \cup \ove{(A \cup B)} = \ove{A \cup (A \cup B)} =
\ove{A \cup B}$, by the same theorem.; \item (vii) The hypothesis
says that $A \sqcup C = \oC$ and $B \sqcup C = \oC$, that is, $\oA
\cup \oC = \oC$, hence $\oA \subseteq \oC$. In the same vein, $\oB
\subseteq \oC$. But these results entail that $\oA \cup \oB
\subseteq \oC$, hence $\ove{(\oA \cup \oB)} \subseteq \ove{\oC}$,
then $\ove{(\oA \cup \oB)} \cup \oC = \oC$; \item (viii)
$\mathbf{0} \sqcup A = \ove{\mathbf{0}} \cup \oA = \oA$, and $A
\sqcup \mathbf{1} = \oA \cup \ove{U} = U = \mathbf{1}$; \item (ix)
If $A \leq B$, then $A \sqcup B = \oB$. But this entails that $\oA
\subseteq \oB$. Thus $\ove{A \cap B} \subseteq \oA \cap \oB = \oA$
[Theorem\ 3.1 (iii)], that is, $A \sqcap B = \oA$. \cqd
\end{enumerate}

\vspace{2mm} Alternatively, we could define $A \leq_1 B$ iff $A
\sqcap B = \oA$. The theorem above follows, with the exception of
item (ix), which should be substituted by $A \leq B \Rightarrow A
\sqcup B = \oB$. Really, assuming this definition, we have $A
\sqcup B = \oA \cup \oB = \oB$, for the hypothesis entails that $A
\sqcap B = \oA$, that is, $\oA = \ove{A \cap B} \subseteq \oA \cap
\oB$ by Theorem\ 3.1 (iii). So $\oA \subseteq \oB$, then $\oA \cup
\oB = \oB$, that is, $A \sqcup B = \oB$. Item (ix) of the theorem
and this result show that $A \leq B$ iff $A \leq_1 B$.

We have proved that $\leq$ is both reflexive and transitive ((i)
and (iii) above), but only ``partially" anti-symmetric, that is,
$A \leq B$ and $B \leq A$ entail $\oA = \oB$. Thus,  $\langle
\mathcal{P}(U), \leq \rangle$ is a kind of ``weak" poset. Since it
contains $\mathbf{0}$ and $\mathbf{1}$ and since any two elements
of $U$ have a supremum (namely, $A \sqcup B$) and an infimum
(namely, $A \sqcap B$). $\mathfrak{I}$ is a ``weak lattice", but
of course it is a lattice stricto sensu if we consider only closed
qsets.

The complement of a qset $A$ relative to the universe $U$ is the
sub-qset of $U$, termed $A^{\bot}$, which has no element
indistinguishable from any element of $A$.

\begin{dfn} [$\mathfrak{I}$-involution, or Generalized
$\mathfrak{I}$-complement] $$A^{\bot} \igual U - \oA.$$  \end{dfn}

\begin{thm} Let $A, B \in \mathcal{P}(U)$. Then:
\begin{enumerate} \item (i) $\emptyset^{\bot} = U$; \item (ii) $U^{\bot}
= \emptyset$; \item (iii) $U - A^\bot = \oA$; \item (iv)
$\ove{A^{\bot}} = A^{\bot} = {\oA}^{\bot}$; \item (v)
${A^{\bot}}^{\bot} = \oA$ \, ($=A$ when $A$ is a closed qset);
\item (vi) $A \leq B \Rightarrow B^\bot \leq A^\bot$.
\end{enumerate} \end{thm} \Proof   \begin{enumerate}
 \item (i) $\emptyset^{\bot} = U - \ove{\emptyset} = U - \emptyset =
 U$;
 \item  (ii) $U^{\bot} = U - \ove{U} = U - U = \emptyset$;
 \item (iii) $U - A^\bot = U - (U - \oA) = \oA$ because they are all
 closed;
 \item (iv) $\ove{A^{\bot}} = \ove{U - \oA} = U - \oA = A^\bot =
 {\oA}^\bot$. Informally speaking, in $U - \oA$ there are no elements
 indiscernible from  the elements of $\oA$ (according to definition of cloud).
 Thus, it is closed, and coincides with ${\oA}^\bot = U - \ove{\oA}$.
 \item (v) ${A^{\bot}}^{\bot} = U - \ove{A^{\bot}} = U - A^\bot = \oA$ \,
 ($= A$ for closed
 qsets);
 \item (vi) $A \leq B \Rightarrow A \sqcup B = \oB$, hence
 $\oA \cup \oB = \oB$ and $\oA \subseteq \oB$. But this implies that
 $U - \oB \subseteq U - \oA$, that is, $B^\bot \subseteq A^\bot$.
 So $B^\bot \cup A^\bot = A^\bot$, then, by Theorem 3.1 (i),
 $\ove{B^\bot} \cup \ove{A^\bot} = \ove{A^\bot}$,
 hence $B^\bot \sqcup A^\bot = \ove{A^\bot}$ or
 $B^\bot \leq A^\bot$. \cqd
\end{enumerate}

\vspace{2mm} Properties (v) and (vi) of the preceding theorem show
that ${}^\bot$ is an involution for closed qsets. For qsets in
general, we shall call it $\mathfrak{I}$-involution, in the spirit
of the above discussion.

\begin{thm}\label{th3.8} If $A, B \in \mathcal{P}(U)$, then:
\begin{enumerate}
 \item (i)  $A \sqcup A^{\bot} = \mathbf{1}$; \item (ii) $A \sqcap A^{\bot}$ =
 $\mathbf{0}$; \item (iii) $A \sqcup (B \sqcap B^{\bot}) = \oA$ ($= A$ for closed qsets);
 \item (iv) $A\sqcap (B \sqcup B^{\bot}) = \oA$ \, ($= A$ for closed qsets);
 \item (v) (De Morgan) $(A \sqcup B)^{\bot} = A^{\bot} \sqcap B^{\bot}$;
 \item (vi) (``Partial" De Morgan) $(A \sqcap B)^{\bot} \subseteq A^{\bot} \sqcup B^{\bot}$
 (equality holds for closed qsets). \end{enumerate}
 \end{thm}
\Proof
    \begin{enumerate}
    \item (i) $A \sqcup A^{\bot}$ = $\oA \cup \ove{A^\bot}$ =
$\oA \cup A^\bot$ = $\oA \cup (U - \oA) = U =
    \mathbf{1}$;
\item (ii) $A \sqcap A^{\bot} = A \sqcap (U - \oA)$ = $\overline{
A \cap (U - \oA)}$ = $\overline{\emptyset}$ = $\mathbf{0}$; \item
(iii) $A \sqcup (B \sqcap B^{\bot}) = A \sqcup \mathbf{0} = \oA
\cup \emptyset = \oA$ \, ($= A$ for closed qsets); \item (iv)
$A\sqcap (B \sqcup B^{\bot})$ = $A \sqcap \textbf{1} = \oA$ \, ($=
A$ for closed qsets); \item (v) $(A \sqcup B)^{\bot} = (\oA \cup
\oB)^\bot = U - \ove{(\oA \cup \oB)}$ = [Th. 3.1 (iv)] $U -
\ove{(A \cup B)}$ = [Th. 3.4 (i)] $U - (\oA \cup \oB) = (U - \oA)
\cap (U - \oB) = $ [Th. 3.1 (ii) and the fact that the involved
qsets are closed] $\ove{(U - \oA) \cap (U - \oB)} = A^\bot \sqcap
B^\bot$; \item (vi) $(A \sqcap B)^{\bot} = U - \ove{(A \sqcap B)}
= U - \ove{\ove{(A \cap B)}} = U - \ove{(A \cap B)} \subseteq $
[Th. 3.1 (iii); equality holds for closed qsets] $\subseteq U -
(\oA \cap \oB) = (U - \oA) \cup (U - \oB) = A^\bot \cup B^\bot = $
[previous theorem (iv)] $\ove{A^\bot} \cup \ove{B^\bot} = A^\bot
\sqcup B^\bot$. \cqd
\end{enumerate}

\vspace{2mm}
 The lattice $\mathfrak{I}$ is $\mathfrak{I}$-orthomodular, that is,
 if $A \leq B$, we have that $\oB = A \sqcup (A \sqcup B^{\bot})^{\bot}$.

 \begin{thm} [$\mathfrak{I}$-orthomodularity] For all $A, B \in \mathcal{P}(U)$:
 $A \leq B \Rightarrow A \sqcup (A \sqcup B^{\bot})^{\bot} =
 \oB$.
 \end{thm}\label{quasiorthomodular}
 \Proof $A \sqcup (A \sqcup B^{\bot})^{\bot}$ = $A \sqcup (B \sqcap A^{\bot})$ =
 $A \sqcup \ove{(B \cap A^{\bot})}$ = $\oA \cup \ove{(B \cap A^\bot)}$ =
 [Th. 3.1 (iv)] $\ove{A \cup (B \cap A^\bot)}$ = $\ove{(A \cup B) \cap (A \cup A^\bot)}$
 = $\ove{(A \cup B) \cap \mathbf{1}} = \ove{A \cup B}$ [Th. 3.1
 (i)] $\oA \cup \oB = A \sqcup B$ = (by the hypothesis) $\oB$. \cqd

 %$A \leq B$ and $A^\bot \leq C$ entail $A \sqcup (B \sqcap C) = (A \sqcup B) \sqcap (A \sqcup C)$.

\begin{dfn} [Orthogonality] Let $A, B \subseteq U$. We say that
$A$ is orthogonal to $B$, and write $A \perp B$, when: $A \perp B
\igual A \leq B^{\bot}$. Furthermore, a collection $S$ of elements
of $\mathcal{P}(U)$ is called pairwise orthogonal iff for any $A,
B \in S$ such that $A \not= B$, it results that
 $A \perp B$. \end{dfn}

\begin{thm} $A \perp B$ iff $A \cap \oB = \emptyset$. \end{thm}
\Proof  If $A \perp B$, then $A \leq B^\bot$, that is, $A \sqcup
B^\bot = \ove{B^\bot}$. Thus, $\oA \cup \ove{B^\bot} =
\ove{B^\bot}$, so $\oA \subseteq \ove{B^\bot}$, hence $A \subseteq
\ove{B^\bot} = B^\bot$ [by Theorem 3.9 (iv)], so $A \cap \oB =
\emptyset$. Conversely, if $A \cap \oB = \emptyset$, then $A
\subseteq B^\bot$, hence $\oA \cup \ove{B^\bot} = \ove{B^\bot} =
B^\bot$ by Theorem\ 3.9 (iv), that is, $A \perp B$. \cqd

\vspace{2mm} Intuitively speaking, $A \cap \oB = \emptyset$ (by
the way, this could be an alternative definition) says that $A$
has no element indistinguishable from elements of $B$.

In quantum logic, the operations $\leq$ and ${}^{\bot}$ are
usually understood as an \emph{implication relation} and a
\emph{negation relation} respectively. Thus, we may introduce the
concept of \emph{logical incompatibility} just using the idea of
orthogonality (\cite[p.\ 12]{dalgiugre04}): $A$ is incompatible
with $B$ iff $A$ implies the negation of $B$, that is iff they are
orthogonal. The negation of the relation $\perp$ is called
\emph{accessibility} (\emph{ibid.}), written $A \not\perp B$.

All of this show that our structure $\mathfrak{I}$ resembles a
non-distributive orthocomplete orthonormal lattice, and it is a
Boolean lattice if we consider only the closed qsets. Since every
modular ortholattice is orthomodular \cite[p.\ 15]{dalgiugre04},
it is an open question whether our lattice has some similarity
with modular lattices, that is, $A \leq B$ entails $A \sqcup (C
\sqcap B) = (A \sqcup C) \sqcap B$ (we still need to check this
and other results).

%---------------------------------------
\section{Summing up}
We resume here the properties of the quasi-lattice $\mathfrak{I} =
\langle \mathcal{P}(U), \textbf{0}, \textbf{1}, \sqcap, \sqcup,
{}^\bot, \leq \rangle$:
\begin{enumerate}
\item ($\mathfrak{I}$-idempotency) $A \sqcap A = \oA$, $A \sqcup A
= \oA$ \item (Commutativity) $A \sqcap A = B \sqcap A$, $A \sqcup
B = B \sqcup A$ \item (Associativity) $A \sqcap (B \sqcap C) = (A
\sqcap B) \sqcap C$, $A \sqcup (B \sqcup C) = (A \sqcup B) \sqcup
C$ \item ($\mathfrak{I}$-absorption) $A \sqcap (A \sqcup B) =
\oA$, $A \sqcup (A \sqcap B) = \oA$ \item ($\mathfrak{I}$-minimum)
$\mathbf{0} \sqcap A = \mathbf{0}$, $\mathbf{0} \sqcup A = \oA$
\item ($\mathfrak{I}$-maximum) $A \sqcap \mathbf{1} = \oA$, $A
\sqcup \mathbf{1} = \mathbf{1}$ \item ($\mathfrak{I}$-involution -
1) ${A^\bot}^\bot = \oA$ \item ($\mathfrak{I}$-involution - 2) $A
\leq B \Rightarrow B^\bot \leq A^\bot$ \item (Complementation) $A
\sqcap A^\bot = \mathbf{0}$, $A \sqcup A^\bot = \mathbf{1}$ \item
($\mathfrak{I}$-absorption -1) $A \sqcup (B \sqcap B^\bot) = \oA$
\item ($\mathfrak{I}$-absorption-2) $A \sqcap (B \sqcup B^\bot) =
\oA$ \item ($\mathfrak{I}$-De Morgan) $(A \sqcup B)^\bot = A^\bot
\sqcap B^\bot$, $(A \sqcap B)^\bot \subseteq A^\bot \sqcup B^\bot$
\item ($\mathfrak{I}$-orthomodularity) $A \sqcup (A \sqcup
B^\bot)^\bot = \oB$.
\end{enumerate}

As we see, it is a rather unusual mathematical structure which
resembles the non-distributive ortholattice of quantum mechanics.
What the specific $\mathfrak{I}$-properties show is that sometimes
we need to consider the closure of a certain qset for getting the
desired result. If we interpret the qsets of elements of $U$ as
extensions of certain predicates, which might stand for physical
properties, the necessity of considering the closure of the qsets
show that some fuzzy characteristic of these properties are been
shown. In fact, take for instance a qset $A$ as the extension of a
certain property $P$, that is, $A$ should stand for the collection
of objects having the property $P$\footnote{By the way, this is
something that is lacking in the usual discussion on quantum
theories, that is, a right ``semantics", which would enable us to
talk of the extension of the relevant predicates.}. Then, for
instance, if we transform $A$ twice by the operation ${}^\bot$
($\mathfrak{I}$-involution - 1), we do not obtain $A$ anymore, but
the qset of the indiscernible of its elements. It seems that
something is changed when we operate with the collections of
objects of the physical systems: we really \emph{transform} them,
as we really do with quantum systems. But we remark that the
physical interpretation of such a structure and its consequences
is still being investigated. For the moment, let us keep with its
mathematical counterpart only.

%-----------------------------------
\section{The corresponding logic}
In this section, we shall be dealing with the first ideas for an
alternative axiomatization of a logic that has as its algebraic
counterpart the $\mathfrak{I}$-lattice, based on the above
assumptions and definitions. We remark once more that this is only
a preliminary sketch, and maybe some modifications would need to
be done, but let us continue even so. As before, we shall be
working within the theory \Q. The concepts introduced below, which
mirror the standard ones, can be developed in the ``standard part"
of \Q, so that we can use the usual mathematical terminology.
Here, as before, the equality symbol ``=" stands for the
extensional equality of \Q.

\newcommand{\cwed}{\curlywedge}
\newcommand{\cvee}{\curlyvee}
\newcommand{\cto}{\twoheadrightarrow}
\newcommand{\cfor}{\bigwedge}
\newcommand{\cexi}{\bigvee}
\newcommand{\cper}{\Downarrow}
\newcommand{\ctop}{\Uparrow}
\newcommand{\ilatt}{$\mathfrak{I}$}

Let us take our algebra $\mathfrak{I} = \langle \mathcal{P}(U),
\textbf{0}, \textbf{1}, \sqcap, \sqcup, {}^\bot \rangle$.  Now we
shall introduce a generalized (or abstract) logic $\mathcal{L} =
\langle F, \mathcal{T}, \cwed, \cvee, \sim, \cto \rangle$ in the
sense of \cite{cos06}, and we shall continue to use use $\to$
$\wedge$, $\vee$, $\neg$, $\forall$ and $\exists$ as
metalinguistic symbols for implication, conjunction, disjunction,
negation, the universal quantifier, and the existential
quantifier, respectively. The elements of the \Q-set $F$ will be
called \emph{formulas}, and denoted by small Greek letters, while
the elements of $\mathcal{T}$ ($\mathcal{T} \subseteq
\mathcal{P}(F)$) are the \emph{theories} of $\mathcal{L}$, and
denoted by uppercase Greek letters (indices can be used in both
cases).

To begin with, let us see how we link such a logic with the
quasi-lattice $\mathfrak{I}$. Suppose that there is a valuation $v
: F \mapsto \mathcal{P}(U)$ such that:
\begin{enumerate}
\item (i) For any $\alpha \in F$, $v(\alpha) \in \mathcal{P}(U)$;
\item (ii) $\cwed$ and $\cvee$ are binary operations on $F$, and
we denote the corresponding images of the pair $\langle \alpha,
\beta \rangle$ respectively by $\alpha \cwed \beta$ and $\alpha
\cvee \beta$. These operations obey the following rules:
    \begin{enumerate}
    \item (a) $v(\alpha \cwed \beta) = v(\alpha) \sqcap v(\beta)$
    \item (b) $v(\alpha \cvee \beta) = v(\alpha) \sqcup v(\beta)$;
    \end{enumerate}

\item (iii) $\sim$ is a mapping from $F$ into $F$, and we define
$v(\sim\alpha) = (v(\alpha))^{\perp}$, for any $\alpha \in F$.
This means that if $v(\alpha) = A$, then $v(\sim\alpha) = U - \oA$
according to the above definitions; \item (iv)  $F \in
\mathcal{T}$, this is the trivial theory; \item (v) If $\{
\Gamma_i \}_{i \in I}$ is a collection of elements of
$\mathcal{T}$, then $\bigcap \Gamma_i \in \mathcal{T}$.
\end{enumerate}

\vspace{2mm} It is clear that this definition is an algebraic
characterization of our logic $\mathcal{L}$ by means of the
lattice $\mathfrak{I}$. Some immediate consequences of this
definition are: $v(\alpha \cwed \sim\alpha) = \mathbf{0}$,
$v(\alpha \cvee \sim\alpha) = \mathbf{1}$, $v(\sim\sim\alpha) =
v(\alpha)$, etc.

It is well known that in standard quantum logics there is an
``implication-problem", to use Dalla Chiara \emph{et al}.'s words
\cite[p. 164]{dalgiugre04}. That is, all conditional connectives
``that can be reasonably introduced" in quantum logics are
``anomalous" (\emph{ibid.}), and this was taken by some authors as
a motive to criticize quantum logics as not being ``real logics".
As Dalla Chiara \emph{et al.}\ say, there are some conditions that
a conditional would satisfy to be classified as an
implication\footnote{Really, several ``quantum implications" can
be defined, as shown in \cite{mal90}, \cite{megpav03},
\cite{rom06}, but we shall not continue with this discussion here.
One of the first works (to our knowledge) that proposed an
axiomatization of the lattice of quantum mechanics is
\cite{kot67}, in which other conditionals are defined. We had no
access to this paper, but know it from  indirect sources, namely,
\cite{dri75} and \cite{san80}.}. These conditions are:

\vspace{2mm}
\paragraph*{Conditions for an Implication}
\begin{enumerate}
\item (i) identity, that is, $\alpha \stackrel{*}{\to} \alpha$,
being $\stackrel{*}{\to}$ the considered conditional; \item (ii)
\emph{modus ponens}, that is, is $\alpha$ is true and $\alpha
\stackrel{*}{\to} \beta$ is true, then $\beta$ is true (\emph{op.
cit.}, p. 164); \item (iii) In an algebraic semantics, a
sufficient condition is: for any structure $\mathcal{A} = \langle
A, v \rangle$, $\mathcal{A} \models \alpha \stackrel{*}{\to}
\beta$ iff $v(\alpha) \leq v(\beta)$.
\end{enumerate}

We say that a formula $\alpha$ is true in the structure
$\mathfrak{I}$, and write $\mathfrak{I} \models \alpha$ iff
$v(\alpha) = \mathbf{1}$, for any valuation $v$. In this case,
$\mathfrak{I}$ is a model of $\alpha$. We write $\Gamma \models
\alpha$ to mean that every model of (the formulas of) $\Gamma$  is
model of $\alpha$. Finally, $\alpha$ is valid iff it is true in
every structure which is an $\mathfrak{I}$-lattice. In this case,
we write $\models \alpha$. It is quite obvious that our aim is to
prove a completeness theorem for our logic relative to the given
semantic, but to do so we need to introduce the concept of
deduction from a set of premises. To begin the issue we shall
finish only in a forthcoming paper, let us define implication.

\begin{dfn}[$\mathfrak{I}$-conditional] $\alpha \cto \beta \igual \beta
\cvee (\sim\alpha \cwed \sim\beta)$ \end{dfn}

This conditional is quite similar to that one called ``Dishkant
implication'' in \cite{megpav03}. Using the above definitions and
Theorem 3.1 (ii), it is immediate to see that $v(\alpha \cto
\beta) = \ove{v(\beta) \cup (v(\alpha)^\bot \cap v(\beta)^\bot)}$.
Thus,
$$v(\alpha \cto \alpha) =  \ove{v(\alpha) \cup (v(\alpha)^\bot \cap
v(\alpha)^\bot)} = \ove{v(\alpha) \cup v(\alpha)^\bot} =
\ove{\mathbf{1}} = \mathbf{1},$$ \noindent for $\mathbf{1} = U$ is
closed. So, $\models \alpha \cto \alpha$. Furthermore, if
$v(\alpha) = \mathbf{1}$ and $v(\alpha \cto \beta) = \mathbf{1}$,
then $\ove{v(\beta) \cup (v(\alpha)^\bot \cap v(\beta)^\bot)} =
\mathbf{1}$ and, since $v(\alpha) = \mathbf{1}$, we get that
$v(\beta) = \mathbf{1}$. Thus, our conditional obeys conditions
(i) and  (ii) of the Conditions for an Implication. In addition,
we can see that condition (iii) is also fulfilled. In fact, by the
hypothesis, we have $\models \alpha \cto \beta$, so $v(\beta \cvee
(\sim\alpha \cwed \sim\beta)) =  \mathbf{1}$. Call $v(\alpha) = A$
and $v(\beta) = B$. Then $B \sqcup \ove{(A^\bot \sqcap B^\bot)} =
U$, that is, $\oB \cup \ove{(A^\bot \cap B^\bot)} = U$. For this
equality to hold, we need that $\ove{(A^\bot \cap B^\bot)} =
{\oB}^\bot = \ove{B^\bot}$. Then $A^\bot \cap B^\bot = B^\bot$, so
$B^\bot \subseteq A^\bot$, that is (for \Q-sets), $A \subseteq B$.
By Theorem 3.9 (vi), $A \leq B$, that is, $v(\alpha) \leq
v(\beta)$.

Let us make a further remark on this definition. We say that $A
\in \mathcal{P}(U)$ is \emph{definable} by a formula $\alpha \in
F$ if $v(\alpha) = A$. Let $\beta$ be such that $v(\beta) = \oA$.
Is there such a $\beta$? The answer is in the affirmative. Since
$A \subseteq \oA$, then $v(\alpha) \leq v(\beta)$, hence by
condition (iii) above, $\models \alpha \cto \beta$. So, $\beta$ is
any formula implied by $\alpha$. This affirmative makes sense, for
$v(\alpha \cto \beta) = \mathbf{1}$ and $v(\beta) = \oA$ say that
$\oA \sqcup (A^\bot  \sqcap \ove{A^\bot}) = U$, that is,
$\ove{\oA} \cup \ove{(A^\bot \cap \ove{A^\bot})} = A \cup
(\ove{A^\bot}  \cap \ove{A^\bot}) = \oA \cup {\oA}^\bot = U$. This
fact will be important for the definition of the connectives of
the logic $\mathcal{L}$. Finally, let us say that $\alpha
\twoheadleftarrow\!\!\cto \beta \igual (\alpha \cto \beta) \cwed
(\beta \cto \alpha)$.

Next we introduce the notion of syntactical consequence  from a
set of premises, written $\Gamma \vdash \alpha$, as follows, where
$v(\Gamma) = \bigcup[v(\alpha) : \alpha \in \Gamma]$ (the
terminology is from \Q\ -- see again Section 2, if necessary).

\begin{dfn}[Syntactical Consequence] $\Gamma \vdash \alpha$ iff
any theory containing $\Gamma$ (really, the formulas of $\Gamma$)
contains $\alpha$. \end{dfn}

Let $\vdash \alpha$ abbreviates $\emptyset \vdash \alpha$, while
$\alpha \vdash \beta$ abbreviates $\{\alpha\} \vdash \beta$
(recall that they are \Q-sets, so the standard notation can be
used), and $\Gamma \not\vdash \alpha$ says that it is not the case
that $\Gamma \vdash \alpha$. It is immediate to prove the
following theorem:

\begin{thm} In $\mathcal{L}$, we have
\begin{enumerate}
\item (i) $\alpha \in \Gamma \Rightarrow \Gamma \vdash \alpha$. In
particular, $\alpha \vdash \alpha$; \item (ii) $\Gamma \vdash
\alpha \Rightarrow \Gamma \cup \Delta \vdash \alpha$; \item (iii)
If $\Gamma \vdash \alpha$ and for every $\beta \in \Gamma$, we
have that $\Delta \vdash \beta$, then $\Delta \vdash \alpha$;
\item (iv) If $\{\Gamma_i\}_{i \in I}$ is a family of subqsets of
$F$ such that for every $\alpha$, $\alpha \in \Gamma \lra \Gamma_i
\vdash \alpha$, then $\forall \alpha (\alpha \in \bigcap_{i \in I}
\Gamma_i \lra \bigcap_{i \in I} \Gamma_i \vdash \alpha)$.
\end{enumerate}
\end{thm}
\Proof Immediate, for the definition of consequence is standard
(see \cite{cos06}, \cite{men97}). \cqd

\vspace{2mm} We shall not continue to develop the syntactical
aspects of this logic (in algebraic terms, but see \cite{cos06}),
but just try to link it with the semantic aspects sketched above.
The least theory containig $\alpha$ is denoted $T_{\alpha}$, and
it coincides with the intersection of all theories containing
$\alpha$ (\emph{op. cit.}). Thus, $\Gamma \vdash \alpha$ iff
$v(\alpha) \subseteq v(T_{\alpha})$. In particular, if $\Gamma$ is
a theory, that is, $\mathrm{Cn}(\Gamma) \igual [ \alpha : \Gamma
\vdash \alpha] = \Gamma$, then $v(\alpha) \subseteq v(\Gamma)$,
and in particular $v(\alpha) \subseteq \ove{v(\Gamma)}$. Finally,
let us recall that since no deduction theorem holds in quantum
logics \cite{mal90}, the same seems to happen here due to the
nature of our implication (but this is still a open problem).

The last point of this paper, which conduces us to another work,
is the question: how to characterize the logic $\mathcal{L}$
axiomatically? We shall follow the approach of generalized logics
in the sense of \cite{cos06}, but not here.

\section*{Acknowledgements}

This work has been sponsored by FAPESP. The article was written
while H{\'e}rcules Feitosa was at UFSC doing post-doctoral
research.

\end{document}